\documentclass[12pt]{article}
\usepackage{amssymb,amsmath}
\usepackage{epsfig}

\begin{document}

\title{A new formalism for calculation of the partition function of single stranded nucleic acids}

\author{Roumen A. Dimitrov \\
University of Sofia, Faculty of Physics,\\ Department of Theoretical Physics, \\
5, James Bouchier Blvd., 1164 Sofia, Bulgaria, \\e-mail: dimitrov@phys.uni-sofia.bg
}


\maketitle

\begin{abstract}
	A new formalism for calculation of the partition function of single stranded nucleic acids is presented.
Secondary structures and the topology of structure elements are the level of resolution that is used. 
The folding model deals with matches, mismatches, symmetric and asymmetric interior loops, stacked pairs in loop 
and dangling end regions, multi-branched loops, bulges and single base stacking that might exist at duplex ends or at the ends of helices. 
Calculations on short and long sequences show, that for short oligonucleotides, a duplex formation often displays 
a two-state transition. However, for longer oligonucleotides, the thermodynamic properties of the single 
self-folding transition affects the transition nature of the duplex formation, resulting in a population of 
intermediate hairpin species in the solution. The role of intermediate hairpin species is analyzed in the case 
when a short oligonucleotides (molecular beacons) have to reliably identify and hybridize to 
accessible nucleotides within their targeted mRNA sequences. It is shown that the enhanced specificity of the molecular beacons 
is a result of their constrained conformational flexibility and the all-or-none mechanism of their hybridization to the target
sequence.
\end{abstract}

\section{Introduction}
Nucleic acids hold great promise as a design medium for the
construction of nanoscale devices with novel mechanical or chemical function \cite{SNC}. Efforts are currently underway in
many laboratories to use DNA and RNA molecules for applications in transport,
switching \cite{GASRRB,BYATAMFSJN,HYXZSNS}, circuitry \cite{MNSDS}, DNA computing \cite{RBNCCJPRLA} and DNA chips 
\cite{DSDLDMRD,SMJGDDSSM}. Conformational switches or diversity of conformations have been proven or
are suspected to be involved in several important processes such as regulation of gene expression, 
translational regulation, mutation and repair, and others \cite{GWMG,SG,GS}. 
During these processes there are several types of interactions
trough a network of RNA-RNA, RNA-DNA, RNA(DNA)-protein, RNA(DNA) self-folding or RNA(DNA)- small
molecular contacts. 

Comparison of short RNAs/DNAs with different base pairs,
loop sequences, bulges, etc.  has yielded an extremely useful
database of thermodynamic parameters from which the stabilities of conformational states of larger 
nucleic acid sequences can be estimated \cite{FRES8301,SUGN8701,HICD8501,PUGJ8901,BLAR7201}. 
The estimation of the thermodynamic parameters is based on
nearest-neighbor approximation for inter-residue interactions of
closest along the sequence nucleotide residues \cite{BORP7401}.

There have been several major improvements in calculation of the
partition function of a single stranded nucleic acids based on McCaskill
algorithm \cite{MCCJ9001,HOFI9401,MATO9601} or estimation of the
free energy based on free energy minimization and the
corresponding sub-ensemble around the minimum free energy
conformation \cite{ZUKM8901,WILA8601,WATM8301,WATM8502,ZUKM8902}.

In this work secondary structures and the topology of structure elements are the level of resolution that is used. 
However, atomic coordinates are also taken into account in the general expressions. 
Unlike proteins \cite{VDAF}, whose secondary structures usually
depend on the global amino acid sequence, DNA/RNA molecules
are currently thought to assemble in a hierarchical manner \cite{BATRT9901,EDRBBMJD,TRSTP}.
The folding can be conceptually partitioned in the two steps of formation 
of the secondary structure and the spatial structure \cite{ITCB}.
As a result DNA/RNA molecules exhibit a modular structure with individual 
structural motifs demonstrating independent characteristics. 

Therefore, investigation of the overall properties of DNA/RNA molecules based on exploration of variety of local 
structural motifs, their interactions and distributions along the sequence needs an appropriate theoretical approaches. 
In particular, this is especially important in a recent increased interest in 
predicting target sites for antisense oligonucleotides in
highly structured DNA/RNA molecules \cite{SWGSMYCR,GBSTALFK,DMMBSFJWDT,SWGNSMYMR,TVJWSF}. 
Because of the economical value and short experimental cycle, antisense technology has been widly accepted as the tool 
to study functions of a gene and to validate drug targets. Antisense oligonucleotides can 
potentially suppress particular gene expression through mechanism such as RNase H-mediated mRNA cleavage, destabilization of the
target mRNA or aberation of translation or splicing. Understanding the conformational constraints and transformation between
different local structural motifs is of great practical importance. Thus, conformational switches of hairpin-shaped oligonucleotide 
primers can be useful for enhancing the specificity of nucleic acid amplification reactions. Interactions between short 
oligonucleotides or small metabolic molecules can lead to conformational switches in the DNA/RNA target molecules 
\cite{MTMPAS,TSCSNB}. These conformational switches can be used for sensing and modulating complex biochemical networks in 
variety of important biological processes \cite{MJWR,GS}. 

Based on such local structural motifs approach in mind, we will use as a starting point our previous work \cite{RADMZ}, 
where we presented a new formalism for hybridization processes between DNA and RNA molecules.
There hybridization accounted only for stacked pairs, interior loops, bulges and, at the
ends, dangling bases. We did not consider stacked pairs in loop and
dangling end regions as well as multi-branch loops. The formalism was applied only to  
short DNA/RNA sequences. Another limitation was that this new formalism was not 
applied for the estimation of the partition function of self-folding. The self-folding of individual 
DNA/RNA molecules was based on free energy minimization and the 
corresponding sub-ensemble around the minimum free energy conformation at each temperature
as given by mfold program by Zuker \cite{ZUKM8902}. This led to some inconsistency in the overall calculations.
For sequences with non-two state transitions the populations of some intermediate species were poorly predicted. 
Recently, using McCaskill algorithm \cite{MCCJ9001, ZUKM0305}, mfold has been updated and now it is able to calculate not only the 
low energy conformations but the ensemble free energy also. It will be interesting in future to compare mfold 
with the formalism developed here.

In this work we present a new formalism for the estimation of the partition function for self-folding.
The formalism use an approach based on the left, right recursion algorithm we have developed for the free energy
calculation of duplexes \cite{RADMZ}. 
All possible conformations of single stranded DNA or RNA sequences in solution are explored. The folding model 
deals with matches, mismatches, symmetric and asymmetric interior loops, stacked pairs in loop and dangling end regions, 
multi-branched loops, bulges and single base stacking that might exist at duplex ends or at the ends of helices. 
Calculations on short and long sequences show, that for short oligonucleotides, a duplex formation often displays 
a two-state transition. However, for longer oligonucleotides, the thermodynamic properties of the single 
self-folding transition affects the transition nature of the duplex formation, resulting in a population of 
intermediate hairpin species in the solution. The advantage of this new formalism is clearly demonstrated 
especially in the case when one need to design relatively short oligonucleotides (molecular beacons) which have to 
reliably identify and hybridize to accessible nucleotides within their targeted mRNA sequences. 
It is shown that the design will enhance the specificity of molecular beacons if they form a stem-and-loop structure
with constrained conformational flexibility and an all-or-none mechanism of their hybridization to the target sequence.

\section{Methods}

\subsection{Recursive calculation }

With increasing of the temperature the overwhelming majority of
the single stranded form conformations tend toward
their corresponding unfolded states. At each temperature there is an ensemble 
of conformational states where each conformation is characterized with the 
fraction of its base pairs and their location along the sequences which are 
melted at that given temperature. Thus along the sequences we have variety of local 
structural motifs characterized by alternating 
loops -single stranded regions- and double stranded regions. The
location and the length of these local structural motifs depend on their relative
Boltzmann statistical weights. In this work we are interested to calculate the partition
functions of the single-stranded forms based on the method
developed for double-stranded forms.

In our previous work (fig.1) \cite{RADMZ}, the polynucleotide sequences of the double-stranded 
forms are described as follows: sequence $1$ is represented by $S_{1}=r_{11}, r_{12},
r_{13}, r_{1i}, r_{1N_{1}}$ and sequence $ 2 $ is
represented by $S_{2}=r_{21}, r_{22}, r_{23}, r_{2j}, r_{2N_{2}}$, 
where $N_{1}$ and $N_{2}$ stand for their corresponding
lengths and $r_{1i}$ and $r_{2j}$ are the space coordinates of the
corresponding nucleotides of sequences $1$ and $2$. The recursion calculation is 
based on the condition that at least there is a two nucleotides along the sequence 
$1$ and sequence $2$ that are in contact $r_{1i}-r_{2j}$ and $1\leq i\leq N_{1}$,
$1\leq j\leq N_{2}$. The sequence enumeration is from the $5^{'} $- to
the $3^{'}$-end of the sequences. The contact $r_{1i}-r_{2j}$ include
an initiation free energy term necessary to bring the two sequences
together $F^{initiation}$. Each nucleotide pair $r_{1i}-r_{2j}$
formally divide the hybridized form $ S_{1}S_{2} $ of the sequences $
1 $ and $ 2 $ in two parts left $ L $ and right $ R $ in such way that
the free energy $ F\left( S_{1}S_{2}\right) $ of $ S_{1}S_{2} $ is a
sum of the free energies of the left $ FL\left( r_{1i},r_{2j}\right)
$and right $ FR\left( r_{1i},r_{2j}\right) $ parts plus the initiation
free energy $ F^{initiation} $ which is assumed to be the same for all
possible pairs $ r_{1i}-r_{2j} $. Thus,

\begin{equation}
F\left( S_{1}S_{2}\right) = {F\!L}\left( r_{1i},r_{2j}\right) + {F\!R}\left(
r_{1i},r_{2j}\right) +F^{initiation}
\end{equation}

This additive property of the energy rules based on nearest neighbor
approximation forms the bases of the recursion calculations of the
partition function $ S_{1}S_{2} $.  The additivity of the free
energy leads to a multiplication of the partition functions of the
left $ {Z\!L} $ and right $ {Z\!R} $ parts \cite{RADMZ}.

\begin{figure}
\begin{center}
\includegraphics{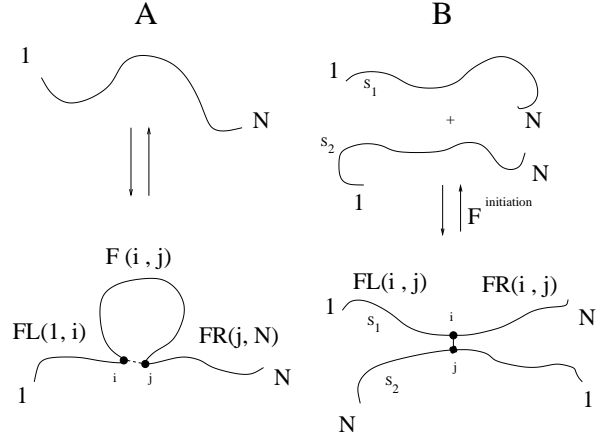}
\caption{Additive property of the free energy rules based on nearest-neighbor approximation: A- self-folding, B- hybridization \cite{RADMZ}.}
\end{center}
\end{figure}

Our main focus in this work is the partition function for single-stranded form which 
similar as we did for the double-stranded form will be described with left and right parts. 
The sequence is represented by $S=r_{1}, r_{2}, r_{3}, \dots, r_{i}, \dots r_{N}$, 
where $N$ stand for it's corresponding length and $r_{i}$ are the space coordinates 
of the corresponding nucleotides of sequences $S$. 
As previously, the recursion calculation is based on the condition that at least 
there is a two nucleotides along the sequence that are in contact $r_{i}-r_{j}$ . 

In contrast to the double-stranded form now the term for the initiation free energy represent the 
formation of a loop between the positions $i$ and $j$ (fig.1).
The sequence enumeration is from the $5^{'} $- to the $3^{'}$-end of the sequence.  
Each nucleotide pair $r_{i}-r_{j}$ formally divide the self-hybridized form  
of the sequences in three parts left $ FL $, middle $ FM $ and right $ FR $ in such way that
the free energy $ F\left( S\right) $ of $ S $ is a
sum of the free energies of the left $ FL\left( r_{i}\right)$, middle 
$ FM\left( r_{i},r_{j}\right) $ and the right $ FR\left( r_{j}\right) $ parts.

\begin{eqnarray}
F\left( S\right) = {F\!L}\left( r_{1},r_{i}\right) + {F\!M}\left(r_{i},r_{j}\right) + {F\!R}\left(r_{j}, r_{N}\right)
\end{eqnarray}

The recursion form of the partition functions of the left, middle and right parts have the forms:

Left part:

\begin{eqnarray}
{Z\!L}\left( r_{1}, r_{i}\right)  & = & {Z\!L}\left( r_{1}, r_{i-1}\right) + \nonumber \\
& & \sum_{1\leq k<i}{Z\!L}\left( r_{1}, r_{k}\right) 
\exp \left( -\frac{{F\!M}\left( r_{k},r_{i}\right)}{RT}\right) \\
{F\!L}\left( r_{1}, r_{i}\right)  & = & -RT \ln \left[ {Z\!L}( r_{1}, r_{i}\right)]
\end{eqnarray}

Middle part:

\begin{eqnarray}
{Z\!M}\left( r_{i},r_{j}\right)  & = & {Z\!M}^{open}\left( r_{i},r_{j}\right) + \nonumber \\
 & & \sum _{i<k<l}\sum _{j>l>k}{Z\!M}\left( r_{k},r_{l}\right)
\exp \left( -\frac{F\left( r_{i},r_{j},r_{k},r_{l}\right)}{RT}\right) 
\end{eqnarray}

\begin{eqnarray}
F\left( r_{i},r_{j},r_{k},r_{l}\right) & = & {F\!L}\left( r_{i},r_{k}\right)+{F\!R}\left( r_{l},r_{j}\right)
\end{eqnarray}

\begin{eqnarray}
{F\!M}\left( r_{i},r_{j}\right)  & = & -RT \ln \left[ {Z\!M}( r_{i},r_{j}\right)]
\end{eqnarray}

Right part:

\begin{eqnarray}
{Z\!R}\left( r_{j}, r_{N}\right)  & = & {Z\!R}\left( r_{j+1}, r_{N}\right) + \nonumber \\
& & \sum_{N\geq k>j}{Z\!R}\left( r_{k}, r_{N}\right) 
\exp \left( -\frac{{F\!M}\left( r_{j},r_{k}\right)}{RT}\right) \\
{F\!R}\left( r_{j}, r_{N}\right)  & = & -RT \ln \left[ {Z\!R}( r_{j}, r_{N}\right)]
\end{eqnarray}

$ {F\!L}\left( r_{1}, r_{i}\right) $ and ${F\!R}\left( r_{j}, r_{N}\right)$  correspond to  
the free energy of self-folding of the $5'$ and $3'$ dangle ends of the sequence.
Obviously, $ {F\!L}\left( r_{1}, r_{N}\right) = {F\!R}\left( r_{1}, r_{N}\right)$. 
The term $ {F\!M}\left(r_{i},r_{j}\right) $ corresponds to the case of initiation of a loop in the middle part. 
Thus, $ {F\!M}^{open}\left(r_{i},r_{j}\right) = -RT\ln[{Z\!M}^{open}\left( r_{i},r_{j}\right)]$ represents the free energy initiation of a loop without internal base pair
contacts. While, $ F\left( r_{i},r_{j},r_{k},r_{l}\right) $ takes into account 
the summation over all possible distribution of structural motifs 
(stack pairs, bulges, symmetric and asymmetric loops, single stranded regions, hairpins and multibranches) 
along the sequences of the interior regions $ (i,k) $ and $ (l,j) $.
For example when $ \left| k-i\right|  = 1 $ and $ \left| l-j\right| =1 $ the free energy $
F\left( r_{i},r_{j},r_{k},r_{l}\right) $ represents a stack pair
which belong to a secondary structure, when $ \left| k-i\right| =2 $
and $ \left| l-j\right| =1 $ or $ \left| k-i\right| =1 $ and $
\left| l-j\right| =2 $ we have a bulge.  When $ \left|
k-i\right| \neq \left| l-j\right| $ and there are no any base pair contacts in the loop regions, 
the free energy $ F\left( r_{i},r_{j},r_{k},r_{l}\right) $ 
represents an asymmetrical internal loop (including the case of a bulge from the one of the
sequences and a loop from the other and another way around), while $
\left| k-i\right| =\left| l-j\right| $ leads to a symmetrical loop
(including the case of a bulge from both sequences). The 
presence of internal base pair contacts in the loop regions lead to hairpins and multibranches.
For detailed description of the free energies of the bulges, symmetric and
asymmetric internal loops and dangling ends we refer the reader to the
recent review by Zuker \cite{ZUKM8904}.

And lastly, based on the multiplication property of the partition functions for the left and  
right parts, for the total partition function we have:

\begin{eqnarray}
Z\left( S\right) =\sum _{1\leq i<j\leq
N}\left[{{Z\!L}\left( r_{1}, r_{i}\right) {Z\!M}\left( r_{i},r_{j}\right){Z\!R}\left( r_{j}, r_{N}\right)}\right]
\end{eqnarray}

\subsubsection{Pair probabilities}

Having calculated the partition function will allow us to derive the
probability distribution of various conformational properties. However, before that we need a  
recursion calculation form for the free energy term $ FL\left( r_{1i},r_{2j}\right)$ in equation
 (1). This term presents the free energy of the left part in case of hybridization. In our 
previous work \cite{RADMZ} we gave an expression for $ FL\left( r_{1i},r_{2j}\right)$ in which 
we did not consider stacked pairs in loop and dangling end regions as well as multi-branch loops. Based on our new formalism
developed above a general recursion calculation form for the left partition function $ {Z\!L}^{h}\left( r_{i},r_{j}\right)$ 
in case of hybridization can be presented as follow:

\begin{eqnarray}
{Z\!L}^{h}\left( r_{i},r_{j}\right)  & = & {Z\!L}\left( r_{1},r_{i}\right){Z\!R}\left( r_{j}, r_{N}\right) + \nonumber \\
 & & \sum _{1\leq k<i}\sum _{N\geq l>j}{Z\!L}^{h}\left( r_{k},r_{l}\right)
\exp \left( -\frac{F\left( r_{i},r_{j},r_{k},r_{l}\right)}{RT}\right) 
\end{eqnarray}

\begin{eqnarray}
{F\!L}^{h}\left( r_{i},r_{j}\right)  & = & -RT \ln \left[ {Z\!L}^{h}\left( r_{i},r_{j}\right)\right]
\end{eqnarray}

 Now we can tern to the calculation of the probabilities of base pairing. For example, the probabilities $ P(r_{i},r_{j}) $ and 
$ P(r_{i},r_{j},r_{i+1},r_{j-1}) $  for single $ r_{i}-r_{j} $ and double $ r_{i}-r_{j},r_{i+1}-r_{j-1}$ 
base pairs are:

\begin{equation}
P\left(r_{i},r_{j}\right) = \frac{{Z\!L}^{h}\left( r_{i},r_{j}\right){Z\!M}\left( r_{i},r_{j}\right)} 
{Z\left( S\right)}
\end{equation}

\begin{equation}
P\left(r_{i},r_{j},r_{i+1},r_{j-1}\right) = \frac{{Z\!L}^{h}\left(r_{i},r_{j}\right)  
{\exp \left( -\frac{F\left(r_{i},r_{j},r_{i+1},r_{j-1}\right) }{RT}\right)}{{Z\!M}\left( r_{i+1},r_{j-1}\right)}} 
{Z\left( S\right)}
\end{equation}

\begin{figure}
\begin{center}
\includegraphics{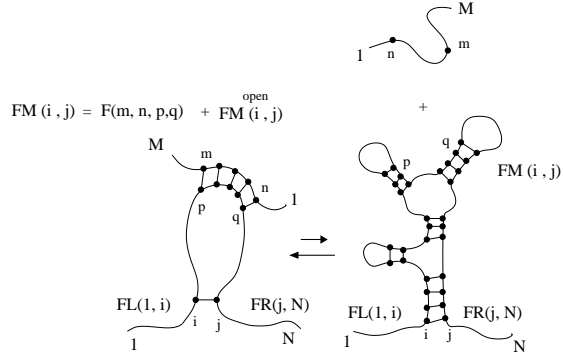}
\caption{Base pair contacts and their free energy contributions in case of an open loop and 
branched hairpin. Also an example is given of conformational switching between the loop and the hairpin as a result 
of interaction of the loop with a short oligo. At the same time the subregion $ \{p,,,q\} $ (involved into a multibranched loop) 
has to unfold before it hybridized with the short oligo.}
\end{center}
\end{figure}

where $ {F\left(r_{i},r_{j},r_{i+1},r_{j-1}\right) }$ is the free energy of base pairing of two nearest-neighbor nucleotides.

Of particular importance is also the ability to monitor 
the transition between the folded and unfolded structures as well
as the partial forms of their conformational intermediates as a function of the temperature by any physical
property that is dependent on the number of base pairs formed.
Fortunately, the absorption spectra as well as thermodynamics are
physical properties that are consistent with the nearest-neighbor
models \cite{PUGJ8901,BLAR7201}.  In other words given
nearest neighbors must have identical values of their absorptions
or melting free energies regardless of their position in the
interior or at the ends of the sequence. In such way the property
monitored as a function of the temperature is proportional to the
fraction of base pairs that are stacked as a nucleic acid molecule
is melted \cite{RADMZ}.

Using the base pairing probabilities we can express the equilibrium fraction of bases paired $ \theta  $ 
as follow:

\begin{eqnarray}
\theta  & = & \sum _{ij}P(r_{i},r_{j}) 
\end{eqnarray}

To calculate the extinction we should take into account
that it is determined by the contribution of the
melted or mismatch loop regions along the constituent sequences of
the self-folded species \cite{CRCIT}. At each given
temperature there is an ensemble of conformation with a narrow or
broad distribution of such loops. The contribution of each of them
is proportional to its relative Boltzmann statistical weight. It
follows from here that the extinction  $ \epsilon(T) $ for the self-folded species can be represented in the form \cite{RADMZ}:

\begin{equation}
\epsilon (T)=\sum ^{N-1}_{i=1}2(1-P(r_{i})- P(r_{i+1})+
P\left(r_{i}, r_{i+1}\right))\xi (i,i+1)-\sum ^{N-1}_{i=1}(1-P(r_{i}))\xi(i)
\end{equation}

where $1-P(r_{i})- P(r_{i+1}) + P\left(r_{i}, r_{i+1}\right)$ is the probability that two closest along the sequence
nucleotides with positions $i$ and $i+1$ are melted and as a result give a contribution $\xi (i,i+1)$ to the total
absorbance. For the probabilities $ P(r_{i}) $ and $ P\left(r_{i}, r_{i+1}\right) $ we have:

\begin{equation}
P(r_{i}) = \sum_{i>n\geq N}{P\left(r_{i},r_{n}\right)}+\sum_{1\leq n<i}{P\left(r_{n},r_{i}\right)}  \nonumber
\end{equation}

\begin{eqnarray}
P\left(r_{i},r_{i+1}\right) & = & \sum_{i+1<n<m}\sum_{n<m\leq N}{P\left(r_{i},r_{i+1},r_{m},r_{n}\right)} + \nonumber \\
& & \sum_{1\leq
n<m}\sum_{n<m<i}{P\left(r_{i},r_{i+1},r_{m},r_{n}\right)} + \nonumber \\
& &  \sum_{i+1<n\leq N}\sum_{1\leq m<i}{P\left(r_{i},r_{i+1},r_{m},r_{n}\right)} 
\end{eqnarray}

The formalism developed in this work allow also incorporation of several types of intramolecular interactions
trough a network of RNA-RNA, RNA-DNA, RNA(DNA)-protein or RNA(DNA)- small
molecular contacts.  The additional free energy terms depending on
the type of interactions (for example hybridization with short oligos or protein molecules) have to be incorporated into 
the free energy term ${F\!M}\left( r_{i},r_{j}\right) $ (fig.2). 

\section{Results and discussions}

\begin{figure}
\begin{center}
\includegraphics{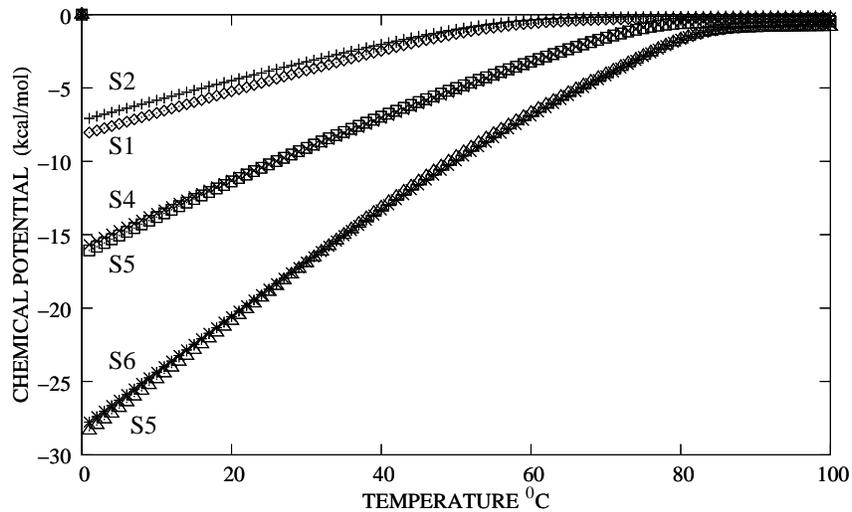}
\caption{Chemical potential versus temperature for the hairpin species formed after dissociation of the three dsDNAs -S1S2, S3S4,
S5S6.}
\end{center}
\end{figure}

Understanding of the molecular forces that control the various sequence- and
solvent-specific conformational forms found within DNA and RNA
oligonucleotides is of great importance. Melting experiments have been the most useful way to 
measure variety of thermodynamic parameters from which the stabilities of larger structures under different conditions can be estimated.
The estimation of the thermodynamic parameters is based on the assumption that the stability of a base pair is
dependent only on the identity of adjacent base pair because the major interactions involved in transformation between different
conformations of the polynucleotide sequence are stacking and hydrogen bonding \cite{SDATD,NSRKDHT,DRHDHT,JDPDHT}.
This additive property of the energy rules based on nearest neighbor
approximation forms the bases of the recursion calculations of the
partition function. The additivity of the free energy leads to a multiplication of the partition functions \cite{RADMZ}.

\begin{figure}
\begin{center}
\includegraphics{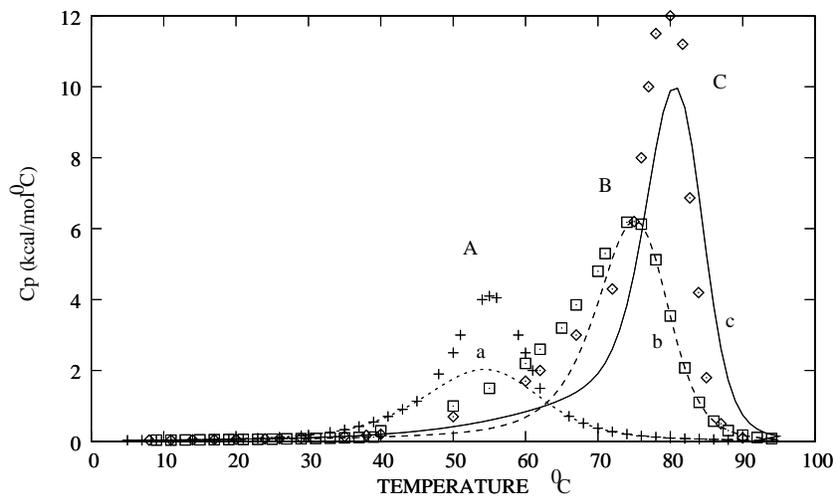}
\caption{Calorimetric excess heat capacity, $\Delta C_{p}$, versus temperature profiles for the three dsDNAs.
Experimental plots for duplex strand transition are as follows \cite{PWNS}: S1S2(A), S3S4 (B), and S5S6 (C). The calculated curves
are with lines and are given as follows: S1S2 (a), S3S4 (b), and S5S6 (c).}
\end{center}
\end{figure}

Based on the multiplication property of the partition function, here we present a new formalism for calculation of the
partition function of a single stranded nucleic acids. The self-folding deal with matches, mismatches, symmetric and asymmetric 
interior loops, bulges and single base stacking that might exist at duplex ends or at the ends of helices. The formalism also takes into 
account base pair contacts in the loop regions,  or dangle ends in the double helix and single hairpin species as well as multi-branches. 
This allow calculations of both short and long sequences.  The self-folding explores all possible conformations of the single strand species.

We did calculations on non-self-complementary DNA sequences with melting temperatures between 
50 $C^o$ and 90 $C^o$. The sequence length is as follows: 9-S1,d(GCTTGTTGC) and S2,d(GCAACAAGC); 
15-S3,d(GCAGGTTGTTTCCGC) and S4,d(GCGGAAACAACCTGC); 21-S5,d(GCAACAGGTTGTTTCCGTTGC) and S6,d(GCAACGGAAACAACCTGTTGC) \cite{PWNS}. 
The self-folding and hybridization between DNA and RNA sequences takes into account the whole ensemble of single and
double strand species in the solution and their fractional extents at different temperatures \cite{RADMZ}.
We assume that the solution can be described as an ensemble of ideally mixed species.
This assumption is based on the experimental evidence that with very good accuracy the single-stranded self-folding trasition 
and the double-stranded association are independent transition processes and the thermodynamic properties and transition 
characteristics of each transition in a mixing solution are identical to those in the isolated systems \cite{PWNS}. The calculated 
chemical potentials of intermadiate hairpin species show that for short oligonucleotides (S1, S2 -fig.3), there is a small thermodynamic 
contribution of the single-strand self-folding transition to the entire transition. As a result the duplex formation for short oligonucleotides
shows a perfectly symmetric two-state shape for the calorimetric excess heat capacity curve versus temperature (fig.4). However, for longer oligonucleotides (S3, S4, S5, S6 -fig.3), calculated chemical potentials 
show that the thermodynamic properties of the single self-folding transition affect the transition nature of the duplex formation, resulting 
in a population of intermediate hairpin species in the solution. The deviation of calculated calorimetric excess heat capacity curves versus 
temperature from a perfectly symmetric shape can be seen for duplexes S3S4 and S5S6 in fig.4. Here, the melting of the intermadiate 
hairpin species are superimposed on the melting of duplex species thus leading to deviation from the two-state shape of the heat capacity curve.

\begin{figure}
\begin{center}
\includegraphics{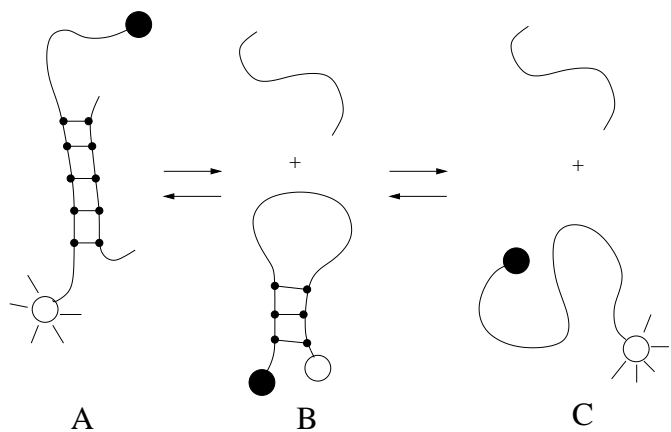}
\caption{Schematic representation of the phase transitions in solutions containing molecular beacons.
At low temperature (phase A) molecular beacons and their targets spontaneously form duplexes. In this 
state molecular beacons are open and fluorescent. At higher temperature (phase B) duplexes 
are destabilized and molecular beacons are released, returning to their closed hairpin conformation, and fluorescence
decreases. As the temperature is raised further (phase C), the closed molecular beacons melt into fluorescent random coils.}
\end{center}
\end{figure}

Further we will analyze in details the transition nature of the duplex formation or dissociation and the role 
of the intermediate hairpin species. 
The role of hairpin intermediates during dissociation or formation of the duplex species in the solution 
is of great importance in the case when a short oligonucleotides 
(molecular beacons) have to reliably identify and hybridize to accessible nucleotides within their targeted mRNA sequences. 
Molecular beacons are DNA probes that form a stem-and-loop intermediate structure and possess an internally
quenched fluorophore. When they bind to complementary nucleic acids, they undergo a conformational transition that 
switches on their fluorescence. Molecular beacons are commonly used to identify complementary strands in the presence of 
unrelated nucleic acids. Understanding the thermodynamic basis and the underlying conformational transformations of the 
enhanced specificity of molecular beacons to their target sequences is of great importance. A simple picture based on 
detailed thermodynamic analysis of the underlying phase transitions in solutions containing molecular beacons is given in fig. 4 \cite{GBSTALFK}. 
Experimental data give evidence for there phases: phase A- probe-target duplex; phase B- free of target molecular beacon in the form of stem-loop 
structure and coiled target; and phase C- molecular beacon and the target are both coiled. All-or-none mechanism is supposed for the 
transitions between the phases. To understand the basis of the molecular beacon specificity from first principle we apply our formalism 
to calculate variety of thermodynamic characteristics such as free energy, enthalpy and entropy. The idea was to compare the behavior 
of molecular beacons in the presence of perfectly complementary target oligonucleotides to their behavior in the presence of targets
whose sequence created a single mismatched base pair in the probe-target duplex. The sequence of the molecular beacon 
used in this work is CGCTCCCAAAAAAAAAAACCGAGCG, and the complementary target GGTTTTTTTTTTTGG. 
In our calculations we do not restrict our self to the case of a two-state transitions where in solution during the
temperature screening there are only two type of conformational species- fully folded and fully unfolded. Rather we consider the
ensemble of all possible intermediate states thus having the most detailed possible picture of the melting process between the
folded and unfolded states of the single and double stranded forms.
Results from our calculations together with the experimental data are given in Table 1. Our calculations are in very good agreement 
with the experimental data \cite{GBSTALFK}. Analysis of the calculated melting curves and intermediates, reveals that the enhanced 
specificity of the molecular beacons is a result of their constrained conformational flexibility and the all-or-none mechanism of 
their hybridization to the target sequence. 

\begin{table}
\caption{Standard enthalpies and standard entropies are shown for solutions containing 50 nM molecular beacons and
1 M target oligonucleotides in the presence of 100 mM KCl and 1 mM $ MgCl_{2}$ \cite{GBSTALFK}. Melting temperatures are for solutions with
50 nM molecular beacons and 300 nM target oligonucleotides. Experiments are given for different mismatches at the 
same position (marked with 0) and the same mismatch at nearest left (marked with -1) and rigth (marked with +1) positions.} 
\fontsize{9}{10pt}\selectfont
\begin{tabular}{|l|c|c|c|c|c|c|c|c|c|c|c|}
\hline {Mismatch}&{Position}&\multicolumn{2}{|c|}{$ -\Delta H^0(kcal/mol)$}&\multicolumn{2}{|c|}{$-\Delta S^0(eu)$}&
\multicolumn{2}{|c|}{$T_m(C^0)$}\\
\hline     &   & exp  & cal  & exp  & cal  & exp & cal\\
           &   &   &   &   &   &  & \\

\hline T-A & 0 & 84  & 80  & 237  & 238  & 42 & 42 \\
\hline A-A & 0 & 69  & 62  & 201  & 202  & 27 & 28\\
\hline C-A & 0 & 61  & 61.2  & 175  & 202  & 23 & 28\\
\hline G-A & 0 & 65  & 61  & 185  & 202  & 28 & 28\\

\hline G-A &-1 & 72  & 65  & 208  & 218  & 29 & 27\\
\hline G-A & 1 & 74  & 65  & 213  & 217  & 29 & 27\\
\hline 
\end{tabular}
\end{table}

Thus, calculations show that the main contribution to the 
free energy of phase A, in case of perfect match between the probe-target sequences, is practically represented by a 
single conformational state of the probe-target duplex. The contributions from bulges, interior loops and dangle ends are 
negligible. The main contributions to 
the free energy of phase B come from the entropy of the coiled target and the free energy of the loop-stem structure 
of the molecular beacon. Flexibility of molecular beacon around its hairpin structure is the main way 
to modulate the stability of phase B. Long stems increase the difference between the melting temperatures of perfectly 
complementary duplexes and mismatched duplexes. However, too long stems make the hairpin 
stable not only in phase B but also in phase A. On the 
other hand, too long hairpin loops decrease the stability of the hairpin.  This can lead to disappearance of phase B. 
Moreover, as the length of the molecular beacon increase, the free energy penalty resulting 
from a mismatched base pair in the probe-target duplex becomes negligible and will decrease the sensitivity to the presence of 
a mismatch. Finally, the free energy of phase C is a sum of the entropies of the random coils of both molecular beacon and its target. 
Our calculations are in full agreement with the experimental data and their thermodynamic analysis (fig. 5)\cite{GBSTALFK}.

\begin{figure}
\begin{center}
\includegraphics{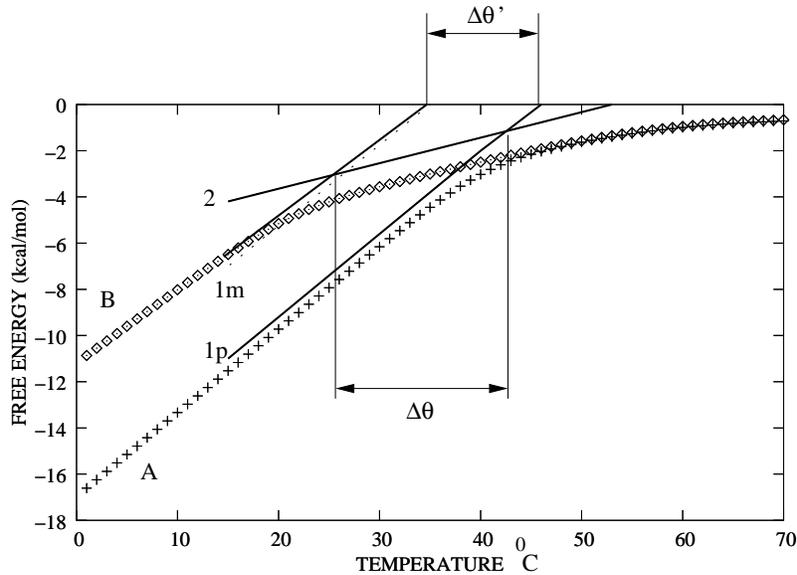}
\caption{Experimental and calculated free energy of a solution of molecular beacons in equilibrium with target oligonucleotides. Experimental
plots \cite{GBSTALFK} for the free energies are as follows: 1p -free energy of the perfect duplex match (phase A); 1m -free energy 
of the mismatch duplex (phase A); 2 - free energy of the molecular beacon closed form and the coiled target (phase B). 
The calculated free energy curves are given as follows: A -free energy of the perfect duplex match (phase A); B -free energy 
of the mismatch duplex (phase A). Since molecular beacons are conformationally more constrained than the unstructured probes, 
line 2 cross the lines 1p and 1m in such way that increase the difference between the melting temperatures of perfectly complementary 
duplexes and mismatched duplexes $\Delta \theta $ compare with the $\Delta \theta^{'}$ for an intermediate state of unstructured probe 
and target.}
\end{center}
\end{figure}

In conclusion, we presented here a general statistical
mechanical approach appropriate to describe the self-folding and hybridization processes of DNA and RNA sequences. 
The folding model deals with matches, mismatches, symmetric and asymmetric interior loops, stacked pairs in loop 
and dangling end regions, multi-branched loops, bulges and single base stacking that might exist at duplex ends or at the ends of helices. 
This allow calculations of both short and long sequences.

Calculations on short and long sequences show, that for short oligonucleotides, a duplex formation often displays 
a two-state transition. However, for longer oligonucleotides, the thermodynamic properties of the single 
self-folding transition affects the transition nature of the duplex formation, resulting in a population of 
intermediate hairpin species in the solution. The advantage of this new formalism is clearly demonstrated 
especially in the case when one need to design relatively short oligonucleotides (molecular beacons) which have to 
reliably identify and hybridize to accessible nucleotides within their targeted mRNA sequences. 
It is shown that the design will enhance the specificity of molecular beacons if they form a stem-and-loop structure
with constrained conformational flexibility and an all-or-none mechanism of their hybridization to the target sequence.
In recent years, a class of diverse regulatory RNAs ( often denoted riboregulators) has emerged that regulate expression
at the posttranscriptional level. These regulatory RNAs fine tune cellular responses to stress conditions, integrating
environmental signals into global regulation. It seems that the structural constraints that enhance the specificity of molecular 
recognition are also a general feature of the mechanism of action of riboregulators. Thus, the formalism developed in this work 
can serve as a first step toward creation of a general approach, which can take into account both affinity and specificity 
of several types of intramolecular interactions trough a network of RNA-RNA, RNA-DNA, RNA(DNA)-protein 
or RNA(DNA)- small molecular contacts.

\bibliography{hybrid}

\end{document}